\documentclass[aps,epsf,twocolumn,showpacs]{revtex4}
\usepackage{amsmath}
\usepackage{epsfig}

\begin{document}

\title{Universality aspects of the $d=3$ random-bond Blume-Capel model}

\author{A. Malakis$^1$}

\author{A. Nihat Berker$^{2,3}$}

\author{N. G. Fytas$^4$}

\author{T. Papakonstantinou$^1$}

\affiliation{$^1$Department of Physics, Section of Solid State
Physics, University of Athens, Panepistimiopolis, GR 15784
Zografos, Athens, Greece}

\affiliation{$^2$Faculty of Engineering and Natural Sciences,
Sabanc$\imath$ University, Orhanl$\imath$, Tuzla 34956, Istanbul,
Turkey}

\affiliation{$^3$Department of Physics, Massachusetts Institute of
Technology, Cambridge, Massachusetts 02139, U.S.A.}

\affiliation{$^4$Departamento de F\'{i}sica Te\'{o}rica I,
Universidad Complutense, E-28040 Madrid, Spain}

\date{\today}

\begin{abstract}
The effects of bond randomness on the universality aspects of the
simple cubic lattice ferromagnetic Blume-Capel model are
discussed. The system is studied numerically in both its first-
and second-order phase transition regimes by a comprehensive
finite-size scaling analysis. We find that our data for the
second-order phase transition, emerging under random bonds from
the second-order regime of the pure model, are compatible with the
universality class of the 3d random Ising model. Furthermore, we
find evidence that, the second-order transition emerging under
bond randomness from the first-order regime of the pure model,
belongs to a new and distinctive universality class. The first
finding reinforces the scenario of a single universality class for
the 3d Ising model with the three well-known types of quenched
uncorrelated disorder (bond randomness, site- and bond-dilution).
The second amounts to a strong violation of universality principle
of critical phenomena. For this case of the ex-first-order 3d
Blume-Capel model, we find sharp differences from the critical
behaviors, emerging under randomness, in the cases of the
ex-first-order transitions of the corresponding weak and strong
first-order transitions in the 3d three-state and four-state Potts
models.
\end{abstract}

 \maketitle

\section{Introduction}
\label{sec:1}

The effect of quenched randomness on the equilibrium and dynamic
properties of macroscopic systems is a subject of great
theoretical and practical interest. It has been known that
quenched bond randomness may or may not modify the critical
exponents of second-order phase transitions, based on the Harris
criterion~\cite{harris74,berker90}. It was more recently
established that quenched bond randomness always affects
first-order phase transitions by conversion to second-order phase
transitions, for infinitesimal randomness in
$d=2$~\cite{aizenman89,hui89} and after a threshold amount of
randomness in $d>2$~\cite{hui89}, as also inferred by general
arguments~\cite{berker93}. A physically attractive description for
these effects has been provided by the Cardy and
Jacobsen~\cite{Cardy97} mapping between the random-field Ising
model and the large $q$ ($d=2$) $q$-state Potts model and the
conjectured extensions of this to higher dimensions. The first
numerical verification of the Cardy-Jacobsen conjecture has been
presented recently in Ref.~\cite{Ferna12}. Furthermore, this
rounding effect of first-order transitions has now been rigorously
established in a unified way in low dimensions ($d\leq 2$)
including a large variety of types of randomness in classical and
quantum spin systems~\cite{greenblatt09}. These
predictions~\cite{aizenman89,hui89} have been confirmed by Monte
Carlo (MC) simulations~\cite{chen92}. Moreover,
renormalization-group calculations on tricritical systems have
revealed that not only first-order transitions are converted to
second-order transitions, but the latter are controlled by a
distinctive strong-coupling fixed point~\cite{falicov96}.

Recently the present authors~\cite{malakis09} have studied details
of the above mentioned expectations for the effects of quenched
bond randomness on second- and first-order phase transitions by
considering the 2d random-bond version of the Blume-Capel (BC)
model. Our studies have provided strong numerical evidence, using
an efficient implementation of the Wang-Landau
algorithm~\cite{wang-01}, clarifying these effects in 2d systems.
In particular it was shown that, the second-order phase
transition, emerging under random bonds from the second-order
regime of the pure 2d BC model, has the same values of critical
exponents as the 2d Ising universality class, with the effect of
the bond disorder on the specific heat being well described by
double-logarithmic corrections, our findings thus supporting the
marginal irrelevance of quenched bond randomness. Furthermore, the
second-order transition, emerging under bond randomness from the
first-order regime of the pure 2d BC model, was shown to belong to
a distinctive universality class with $\nu=1.30(6)$ and
$\beta/\nu=0.128(5)$. These results pointed to the existence of a
strong violation of universality principle of critical phenomena,
since these two second-order transitions, with different sets of
critical exponents, are between the same ferromagnetic and
paramagnetic phases.

In the current MC study, yielding also very accurate information,
we wish to continue our investigations on the 3d BC model. We
consider again a bimodal form of quenched bond randomness and we
find different critical behaviors of the second-order phase
transitions, emerging from the first- and second-order regimes of
the pure model. Hence, in the present paper we offer additional
evidence for the anticipated strong violation of universality. We
will also constructively compare our findings to the existing
literature on the effects of randomness in 3d systems, for which
the pure versions undergo first- and second-order phase
transitions. In the present MC study, we employ a different
numerical scheme based on cluster algorithms and parallel
tempering (PT) and, as we will discuss in detail, this practice is
a very good alternative for second-order phase transitions in
disordered systems.

The relevant literature concerns mainly the 3d Ising model with
quenched randomness, a clear case of the Harris criterion. The
general random model has been extensively studied using MC
simulations~\cite{landau-80,chow-86,heuer-90,hennecke-93,ball-98,wiseman-98,calabrese-03,HaseP07,berche-02,berche-04,Iva-05,fytas-10,Hase07}
but also field theoretical renormalization group
approaches~\cite{folk-00,pakhnin-00,pelissetto-00}. The diluted
model has been treated in the low-dilution regime by analytical
perturbative renormalization group
methods~\cite{newman-82,jug-83,mayer-89} and a new fixed point,
independent of the dilution, has been found. However, the first
numerical studies suggested a continuous variation of the critical
exponents along the critical line, but it became clear, after the
work of Ref.~\cite{heuer-90}, that the concentration-dependent
critical exponents found in MC simulations are the effective ones,
characterizing the approach to the asymptotic regime. Nowadays,
the extensive numerical investigations of Ballesteros \emph{et
al.}~\cite{ball-98}, Berche \emph{et al.}~\cite{berche-04}, and
Fytas and Theodorakis~\cite{fytas-10} for the site-, bond-diluted,
and random-bond versions of the model, have cleared out the doubts
by presenting concrete evidence that the critical behavior of the
3d Ising model with quenched uncorrelated disorder is controlled
by a new random fixed point, independent of the disorder strength
value and the way this is implemented in the system. Moreover,
both site- and bond-diluted Ising systems have been also studied
by Hasenbusch \emph{et al.}~\cite{HaseP07} in a high-statistics MC
simulation, confirming universality and providing very accurate
estimates of the critical exponents.

However, although the evidence of the existence of a unique
universality class in the 3d random Ising model (RIM) is very
strong, a full characterization of the types of models and the
types of disorder that lead a transition into this class is still
a challenging question. Such an interesting case is the $\pm J$
Ising model at the ferromagnetic-paramagnetic transition line, for
which Hasenbusch \emph{et al.}~\cite{Hase07} have shown that it
belongs also to the RIM universality class. Another relevant 3d
random model is the site-diluted $q=3$ Potts model which, in its
pure version, is known to have a weak first-order phase
transition. This model was studied by Ballesteros \emph{et
al.}~\cite{ball-00} and it was clearly verified that the emerging
under randomness second-order transition gives a definitively
different magnetic exponent $\eta$. An analogous study on the 3d
site-diluted $q=4$ Potts model which, in its pure version, is
known to exhibit a strong first-order phase transition has also
produced results consistent with a further distinctive
universality class~\cite{chate-01}. The present study of the 3d
random-bond BC model gives us the opportunity of a direct
comparison with the above relevant 3d cases, by considering the
emerging second-order transitions stemming from both its
Ising-like continuous and its strong first-order transitions,
making thus our attempt more appealing for understanding better
the effect of disorder in 3d systems. Finally, it is to be noted
that, although there is a number of recent
papers~\cite{Puha-01,Salmon-10,Wu-10,Dias-11} dealing with the
effects of randomness on the 3d ferromagnetic BC model, the
present study appears to be the first attempting to clearly
distinguish between the universality aspects of the ex-second- and
ex-first-order regimes of the 3d random BC model.

The rest of the paper is laid out as follows: In the following
Section we define the model (subsection~\ref{sec:2a}) and we give
a detailed description of our numerical approach
(subsection~\ref{sec:2b}) utilized to derive numerical data for
large ensembles of realizations of the disorder distribution and
lattices with linear sizes within the range $L\in \{8-44\}$. In
subsection~\ref{sec:2b} the finite-size scaling (FSS) scheme is
described. In Sec.~\ref{sec:3} we discuss the effects of disorder
on the second-order phase transition regime of the model.
Respectively, in Sec.~\ref{sec:4} we illustrate the conversion,
under random bonds, of the originally first-order transition of
the model to second-order and we present full details of our FSS
attempts, that strongly point to a new distinctive universality
class. Our conclusions are summarized in Sec.~\ref{sec:4}.

\section{Model, Simulation, and Finite-size scaling schemes}
\label{sec:2}

\subsection{The random-bond Blume-Capel model}
\label{sec:2a}

The pure BC model~\cite{blume66,capel66} is defined by the
Hamiltonian
\begin{equation}
\label{eq:1}
H_{p}=-J\sum_{<ij>}s_{i}s_{j}+\Delta\sum_{i}s_{i}^{2},
\end{equation}
where the spin variables $s_{i}$ take on the values $-1, 0$, or
$+1$, $<ij>$ indicates summation over all nearest-neighbor pairs
of sites, and $J>0$ the ferromagnetic exchange interaction. The
parameter $\Delta$ is known as the crystal-field coupling and to
fix the temperature scale we set $J=1$ and $k_{B}=1$.

This model is of great importance for the theory of phase
transitions and critical phenomena and besides the original
mean-field theory~\cite{blume66,capel66}, has been analyzed by a
variety of approximations and numerical approaches, in both $d=2$
and $d=3$. These include the real space renormalization group, MC
renormalization-group calculations~\cite{landau72},
$\epsilon$-expansion renormalization groups~\cite{stephen73},
high- and low-temperature series calculations~\cite{fox73}, a
phenomenological FSS analysis using a strip
geometry~\cite{nightingale82,beale86}, and of course MC
simulations~\cite{malakis09,jain80,landau81,care93,deserno97,Blote95,Blote04,silva06}.
As mentioned already in the introduction the phase diagram of the
model consists of a segment of continuous Ising-like transitions
at high temperatures and low values of the crystal field which
ends at a tricritical point, where it is joined with a second
segment of first-order transitions between ($\Delta_{t},T_{t}$)
and ($\Delta_{0},T=0$). For the simple cubic lattice, considered
in this paper, $\Delta_{0}=3$. The location of the tricritical
point has been estimated by Deserno~\cite{deserno97} using
microcanonical MC approach as the point
$[\Delta_{t},T_{t}]=[2.84479(30),1.4182(55)]$, and to a better
accuracy by the more recent estimation of Deng and
Bl\"{o}te~\cite{Blote04} as
$[\Delta_{t},T_{t}]=[2.8478(9),1.4019(2)]$.

In the random-bond version of the BC model, studied in this paper,
the bond disorder follows the bimodal distribution
\begin{align}
\label{eq:2}
P(J_{ij})~=~&\frac{1}{2}~[\delta(J_{ij}-J_{1})+\delta(J_{ij}-J_{2})]\;;\\
\nonumber
&\frac{J_{1}+J_{2}}{2}=1\;;\;\;J_{1}>J_{2}>0\;;\;\;r=\frac{J_{2}}{J_{1}}\;,
\end{align}
where the parameter $r$ reflects the strength of the bond
randomness (ratio of interaction strengths in a fixed $50\%/50\%$
weak/strong bonds mixing). The resulting quenched disordered
version of the model reads now as
\begin{equation}
\label{eq:3} H=-\sum_{<ij>}J_{ij}s_{i}s_{j}+\Delta
\sum_{i}s_{i}^{2}.
\end{equation}

Since we aim at investigating the effects of bond randomness on
the segments of continuous Ising like and first-order transitions
of the original pure model, we have to choose suitable values of
the crystal field $\Delta$ and the disorder parameter $r$. For the
first-order regime, we are looking for a convenient strong
disorder combination, strong enough to convert the originally
first-order transition to a second-order one. The value
$\Delta=2.9$ is a convenient choice, since it is well inside the
region of first-order transitions (see the above mentioned
location for the tricritical point). At this value of $\Delta$ we
have numerically verified that the disorder strength
$r=0.5/1.5=1/3$ is strong enough to convert, without doubt, the
original first-order transition to a genuine second-order
transition. Thus, we will consider, in the sequel, the cases
$(\Delta,r)=(1,1/3)$ and $(\Delta,r)=(2.9,1/3)$, which correspond
to the ex-second- and ex-first-order regimes. These cases serve
very well our purposes to study the universality aspects of the
emerging under bond-randomness second-order phase transitions of
the model.

\subsection{Reviews of Monte Carlo and finite-size scaling schemes}
\label{sec:2b}

The accuracy of MC data may be decisive for a successful FSS
estimation of critical properties and the proper selection of an
algorithm is the basic requirement for the generation of accurate
MC data. Thus, over the years, the numerical estimation of
critical exponents has been a non-trivial exercise, even for the
simplest models, such as the Ising model. An appropriate algorithm
close to a critical point has to overcome the well-known effects
of critical slowing down, and some otherwise excellent and exact
algorithms, such as the Metropolis algorithm~\cite{metro53}, which
work adequately far from the critical point, become inefficient
close to it. It is also well known that, an approach via
importance sampling, close to a second-order phase transition,
requires appropriate use of cluster algorithms that can
efficiently overcome the critical slowing down effects. Wolff-type
algorithms~\cite{Swendsen87,Newman99,LandBind00} belong to this
category, are easy to implement, and also very efficient close to
critical points.

Since, in the present paper, we wish to simulate the random-bond
BC model close to critical points, the implementation of the Wolff
algorithm could be a suitable alternative. However, for the BC
model the Wolff algorithm can not be used alone, because Wolff
steps act only on the non-zero spin values. A suggested practice
is now a hybrid algorithm along the lines followed by
Ref.~\cite{Blote95}. An elementary MC step (MCS) of this hybrid
scheme consists of a number of Wolff steps (typically 5
Wolff-steps) followed by a Metropolis sweep of the lattice,
whereas the usual MCS for the Metropolis algorithm is just one
lattice sweep. The combination with the Metropolis lattice sweep
is dictated by the fact that the Wolff steps act only on the
non-zero spin values. The proposed hybrid algorithm is, of course,
one of several alternatives for the study of the BC model. We note
in passing that, in our previous analogous studies of the square
lattice pure and random-bond BC model~\cite{malakis09} we have
been using a sophisticated two-stage Wang-Landau approach, very
efficient for studying at the same time characteristics of the
first- and second-order regimes of the model.

However, since the present study concerns only the behavior close
to the emerging, under bond randomness, critical points, the
described hybrid approach, combined with a PT protocol, which will
be detailed below, is much more appealing and its convergence may
be easily checked. Thus, we have simulated the random-bond BC
model on the simple cubic lattice for the cases
$(\Delta,r)=(1,1/3)$ and $(\Delta,r)=(2.9,1/3)$ and for lattice
sizes in the range $L=8-44$, by implementing the hybrid approach
described above, suitably adapted to the present disordered model.
As is always the case, the simulation task for a disordered
system, such as the present random-bond BC model, is by far more
demanding than its pure counterpart, since one has to sum over
disorder realizations. Furthermore, for the random-bond model the
simple Wolff algorithm cannot be applied and a suitable
generalization is necessary. This generalization is a
straightforward adaption of the Wolff algorithm by introducing two
different probabilities for putting links between sites;
corresponding to the weak and strong nearest-neighbor
ferromagnetic interactions. It is then rather easy to show that,
the generalized cluster algorithm has all the properties of the
original Wolff algorithm.

In order to extract information on the critical properties of the
random-bond BC model, we are going to apply the ideas of FSS. For
the estimation of the critical temperatures and the corresponding
critical exponents, one has to generate MC data to cover several
finite-size anomalies of the finite systems of linear size $L$.
\begin{figure}[htbp]
\includegraphics*[width=9 cm]{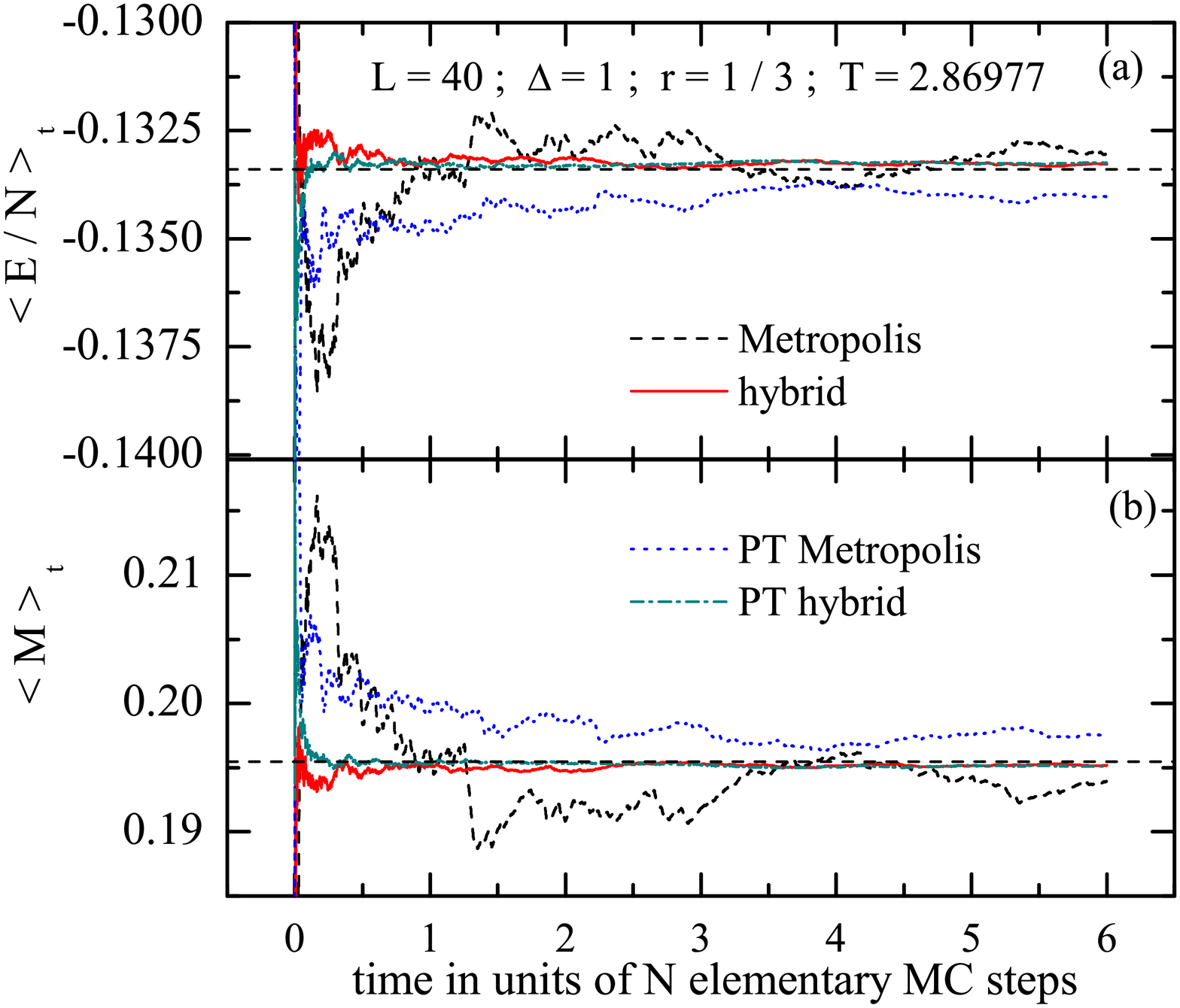}
\caption{\label{fig:1} (color online) Behavior of moving averages
for the energy per site ${\langle E/N\rangle}_{t}$, in panel (a),
and the order parameter ${\langle M\rangle}_{t}$, in panel (b), at
a selected temperature, close to the critical. The behavior of the
four simulation schemes is illustrated and the definition of
elementary steps is discussed in the text.}
\end{figure}
The above described hybrid approach has to be carried over to a
certain temperature range depending on the lattice size. These
temperatures may be selected independently or it may also be
convenient, and accurate, to implement a PT protocol. Such a PT
protocol, based (mainly) on temperature sequences corresponding to
an exchange rate $0.5$, was tested for both cases considered in
this work, i.e. $(\Delta,r)=(1,1/3)$ and $(\Delta,r)=(2.9,1/3)$,
and was found to be very accurate. We should point out here that,
the proposed PT approach is very close to the practice suggested
recently in Ref.~\cite{bittner11}. Furthermore, a similar approach
has been followed recently for a very successful estimation of
critical properties of the Ising and the $S=1$ model (BC) on some
Archimedean lattices~\cite{malakis12}. Appropriate temperature
sequences, for the application of a hybrid PT protocol were
generated via short preliminary runs. During these preliminary
runs a simple histogram method~\cite{Swendsen87,Newman99} was
applied to yield energy probability density functions and from
these, the appropriate sequences of temperatures were easily
determined, following the recipe of Ref.~\cite{bittner11}. The
preliminary runs were applied to only one disorder realization,
but is was checked that the resulting temperature sequences were
producing, to an excellent approximation, a constant exchange rate
of $0.5$ for the PT protocol, when applied to any other different
disorder realization. This is of course an intrinsic, and very
convenient property of the present random-bond model.

The above MC scheme was carefully tested almost for all lattice
sizes before its implementation for the generation of MC data
necessary for applying FSS. These tests included two basic
ingredients: the first was the estimation of the MC times
necessary for equilibration and thermal averaging process applied
to a particular disorder realization and the second involved the
observation of running sample-averages of some thermodynamic
quantity, such as for instance, the magnetic susceptibility, at
some particular temperature close to the critical one. These
tests, not only provided convincing evidence for a successful
determination of equilibrium results, but also highlighted the way
towards an efficient and optimum scheme. For instance it was
checked that for a rather small lattice size, such as $L=16$, the
sample averages over $150$ disorder realizations of all
thermodynamic quantities were identical, irrespective of the
algorithm used (simple Metropolis or hybrid algorithm), provided
one uses reasonable MC times for equilibration and thermal
averaging. This could be explained as a result of effective
cancellation of statistical errors from the sample-averaging
process. Noteworthy, that sample-to-sample fluctuations are much
larger than any usual statistical errors in a proper MC thermal
averaging process. Of course, for a particular disorder
realization the difference in the outcome of different algorithms
may be truly spectacular.

In fact, we can observe the superiority of the hybrid approach,
over a simple Metropolis scheme~\cite{metro53}, in
Fig.~\ref{fig:1}. This figure is constructed by using moving
averages for the energy per site ($\langle E/N \rangle_{t}$) and
the order parameter ($\langle M \rangle_{t}$) for a lattice size
$L=40$ at a selected temperature, close to the critical
temperature, for the case $(\Delta,r)=(1,1/3)$ of the random-bond
BC model. As it can be seen from this illustration, the Metropolis
algorithm suffers from very strong fluctuations and so does the
corresponding PT protocol (applied in a temperature sequence
including the temperature shown) using this algorithm. They both
follow a very slow approach to equilibrium and only with the help
of extensive sampling one could expect results of reasonable
accuracy. On the other hand, the hybrid approach converges very
fast to equilibrium and produces quite accurate results. The
corresponding PT hybrid algorithm appears to further improve this
good behavior, which is a reasonable expectation, since the PT
approach should be in general more effective for disordered
systems. Note also that, Fig.~\ref{fig:1} is constructed by using
one particular disorder realization and the dashed straight lines
in the panel of the figure indicate the final average values of
four independent PT hybrid runs of the same realization, whereas
the fluctuating lines correspond to one single run for each
algorithm. A careful examination of this illustration shows that
even a single run for the thermal averaging process is very
accurate for the hybrid and the PT hybrid algorithms and for the
presented case, a MC time of the order of $3\times N,$ is adequate
and optimum for the thermal process. The corresponding
equilibration MC times are always shorter than this. There is no
doubt that, the use of the hybrid algorithm is essential for the
study of the present model.

Let us review now details from the FSS tools used throughout the
paper, for the estimation of critical properties of the disordered
system. The outline below follows our previous studies on
random-bond models and is very close to the traditional standard
tools of the theory of FSS. When studying a disordered system, a
large number of disorder realizations has to be used in the
summations in order to obtain good sample-averages of any basic
thermodynamic quantity $Z$, which is a usual thermal average of a
single disorder realization. The sample (or disorder) averages may
be then denoted as $[Z]_{av}$ and their finite-size anomalies
respectively as $[Z]_{av}^{\ast}$. These (disorder-averaged)
finite-size anomalies will be used in our FSS attempts, following
a quite common practice~\cite{chate-01} and their temperature
locations will be denoted by $T_{[Z]_{av}^{\ast}}$. Thus, we also
follow the conventional numerical approach that appears to work
well for the present model, at both the ex-second- and
ex-first-order regimes, at the disorder strength values
considered. However, in an alternative approach~\cite{wiseman-98}
one may consider individual sample dependent maxima (anomalies)
and the corresponding sample dependent pseudocritical
temperatures. This alterative route is far more demanding
computationally, but the corresponding FSS analysis may be more
precise, and additional useful information concerning the
properties of disorder averages becomes available. Our study
addresses only exponents describing the disorder-averaged behavior
and we do not use a FSS analysis based on sample dependent
pseudocritical temperatures. For disordered systems one can make a
clear distinction between typical and averaged
exponents~\cite{DSFish95,Chay86}. An extreme example of this case
with very pronounced differences in the corresponding $\nu$
exponents has been provided, and critically discussed, by
Fisher~\cite{DSFish95} on the random transverse-field Ising chain
model. Finally, disordered critical phenomena are known to display
in general multicritical exponents~\cite{Ludw90,Month09} and we
shall return to this interesting point in our last section, when
discussing the violation of universality in the present model.

The number of disorder realizations and the selection of
temperatures may influence the accuracy and suitability of the MC
data from which the locations of the finite-size anomalies are
determined, by fitting a suitable curve (for instance a
fourth-order polynomial) in the neighborhood of the corresponding
peak. Since we are implementing a PT hybrid approach, based on
temperatures corresponding to an exchange rate $0.5$, we are
selecting certain temperature sequences consisting of a number of
(say 3 or 5) different temperatures, averaging then over a
relatively moderate number ($\sim 100$) of disorder realizations.
This yields a number of (3 or 5) points of the averaged curves
$[Z]_{av}$. This set of points may not be adequately dense and
will not cover the ranges of the peaks of all thermodynamic
quantities of interest. Therefore, this procedure is repeated
several times (depending on the linear size $L$) by using new sets
of temperatures which are translated with respect to the previous
sets. These translations should be carefully chosen so that a
final dense set of points, suitable for all finite-size anomalies,
is obtained. This corresponds, on average, to a very large number
of realizations, since for each PT hybrid run, a different set of
disorder realizations was used. We have found this practice very
convenient, efficient, and most importantly quite accurate. Thus,
we were able to describe the averaged finite-size anomalies of the
system with high accuracy.

In order to estimate the critical temperature, we follow the
practice of simultaneous fitting approach of several
pseudocritical temperatures~\cite{malakis09}. From the MC data,
several pseudocritical temperatures are estimated, corresponding
to finite-size anomalies, and then a simultaneous fitting is
attempted to the expected power-law shift behavior
$T_{[Z]_{av}^{\ast}}=T_{c}+b_{Z}\cdot L^{-1/\nu}$. The
traditionally used specific heat and magnetic susceptibility
peaks, as well as the peaks corresponding to the following
logarithmic derivatives of the powers $n=1$, $2$, and $n=4$ of the
order parameter with respect to the inverse temperature
$K=1/T$~\cite{ferrenberg91},
\begin{equation}
\label{eq:4} \frac{\partial \ln \langle M^{n}\rangle}{\partial
K}=\frac{\langle M^{n}H\rangle}{\langle M^{n}\rangle}-\langle
H\rangle,
\end{equation}
and the peak corresponding to the absolute order-parameter
derivative
\begin{equation}
\label{eq:5} \frac{\partial \langle |M|\rangle}{\partial
K}=\langle |M|H\rangle-\langle |M|\rangle\langle H\rangle,
\end{equation}
will be implemented for a simultaneous fitting attempt of the
corresponding pseudocritical temperatures. Furthermore, the
behavior of the crossing temperatures of the fourth-order Binder's
cumulant~\cite{binder81}, and their asymptotic trend, will be
observed and utilized for a safe estimation of the critical
temperatures.

The above described simultaneous fitting approach provides also an
estimate of the correlation length exponent $\nu$. An alternative
estimation of this exponent is obtained from the behavior of the
maxima of the logarithmic derivatives of the powers $n=1$, $2$,
and $n=4$ of the order parameter with respect to the inverse
temperature, since these scale as $\sim L^{1/\nu}$ with the system
size~\cite{ferrenberg91}. Once the exponent $\nu$ is well
estimated, the behavior of the values of the peaks corresponding
to the absolute order-parameter derivative, which scale as $\sim
L^{(1-\beta)/\nu}$ with the system size~\cite{ferrenberg91}, gives
one route for the estimation of the magnetic exponent ratio
$\beta/\nu$. Additionally, knowing the exact critical temperature,
or very good estimates of it, we can utilize the behavior of the
order parameter at the critical temperature for the traditionally
effective estimation of the exponent ratio $\beta/\nu$
($M_{c}=M(T=T_{c})\sim L^{-\beta/\nu}$). Summarizing, our FSS
approach utilizes, besides the traditionally used specific heat
and magnetic susceptibility maxima, the above four additional
finite-size anomalies for an accurate estimation of critical
temperatures and relevant exponents.

\section{Ex-second-order regime: Universality class of the random Ising model}
\label{sec:3}

We study in this Section the 3d random-bond BC model at the
second-order regime of the phase diagram. We have considered the
case $(\Delta,r)=(1,1/3)$ and simulated lattices with linear sizes
in the range $L = 8 - 44$. Below we discuss in detail our efforts
to find a comprehensive FSS scheme to fit the numerical data.
However, before illustrating details of our study, it is useful to
recall again that, according to general universality arguments,
the pure BC model at $\Delta=1$ is expected to belong to the 3d
Ising universality class and thus, the case studied here, should
be contrasted to the relevant literature on the 3d RIM. From the
relevant literature, it appears that a consensus have been
achieved today with regard to the existence of a single
universality class for the general 3d
RIM~\cite{ball-98,berche-04,fytas-10,Hase07}. However, it is also
true that the study of Berche \emph{et al.}~\cite{berche-04} on
the 3d bond-diluted Ising model, has emphasized the strong
influence of crossover phenomena and pointed out that the
identification of this universality class is a very difficult
task. For the case of magnetic bond concentration $p=0.7$, studied
in that paper, these authors found an effective exponent
$(1/\nu)_{eff}=1.52(2)$. Correspondingly, for the case $p=0.55$
the effective value was found to be $(1/\nu)_{eff}=1.46(2)$. This
noticeable variation in the effective exponents indicated possible
confluent corrections and/or crossover terms~\cite{berche-04}.
They finally agreed that, the case $p=0.55$ appears to be the case
of least scaling corrections and thus, in this way, agreement with
the value given by Ballesteros \emph{et al.}~\cite{ball-98} was
established.

The problem of slowing decaying scaling corrections has been
discussed in detail for both these randomly site- and bond-diluted
systems by Hasenbusch \emph{et al.}~\cite{HaseP07}, and for the
latter, the value $p=0.54(2)$ has been proposed as the value at
which the leading scaling corrections vanish. The above
observations will be very useful later in this Section, when we
discuss our problems in extracting a reliable estimate of this
exponent, since we have faced similar problems. Noteworthy that,
in a systematic study of the $\pm J$ Ising model at the
ferromagnetic-paramagnetic transition line, Hasenbusch \emph{et
al.}~\cite{Hase07} have analyzed the effects of leading and
next-to-leading scaling corrections and proposed a value of the
concentration of the ferromagnetic lines of this model at which
corrections to scaling almost vanish. The final estimate for the
correlation length exponent determined by this study is almost
identical with the value given by Ballesteros \emph{et
al.}~\cite{ball-98}.

\begin{figure}[htbp]
\includegraphics*[width=9 cm]{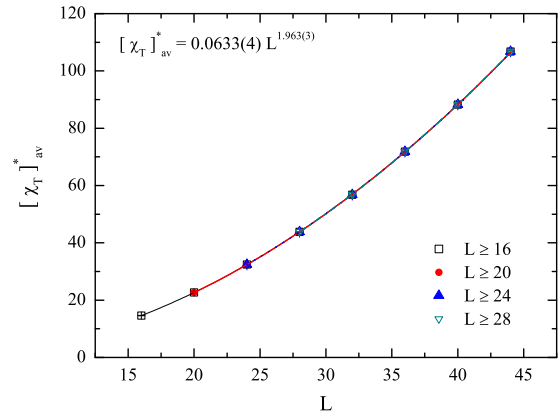}
\caption{\label{fig:2} (color online) Illustration of the
divergence of the susceptibility maxima by a simultaneous global
fitting attempt on the data corresponding to several lattice-size
ranges shown in the panel, indicated by the different line
colors.}
\end{figure}

We start the FSS analysis by attempting the estimation of the
exponent ratio that characterizes the divergence of the
susceptibility. We assume that the (sample-averaged)
susceptibility obeys a simple power law of the form
$[\chi_T]_{av}^{\ast}=b\cdot L^{\gamma/\nu}$. The fitting attempts
are found to be quite stable with respect to the size range
chosen. We attempted the ranges $L=8-44,\;12-44,\;\ldots,\;28-44$
and observed the behavior of the resulting coefficients $b$ and
exponents $\gamma/\nu$. The estimates of the exponent vary slowly
in the range $\gamma/\nu=1.962-1.967$, as we increase $L_{min}$
from $L_{min}=8$ to $L_{min}=28$. A kind of mean value estimate
can be obtained by a global fitting attempt simultaneously applied
to the above ranges. Such a simultaneous attempt using the ranges
$L=16-44$, $20-44$, $24-44$, and $L=28-44$ is shown in
Fig.~\ref{fig:2} giving an estimate $\gamma/\nu=1.963(3)$. The
illustrated estimate is in full agreement with the values given by
Ballesteros \emph{et al.}~\cite{ball-98} and Berche \emph{et
al.}~\cite{berche-04}. Finally, observing the over all trend of
such simultaneous fitting attempts we shall propose as our final
estimate $\gamma/\nu=1.964(4)$, in coincidence with the estimate
$\gamma/\nu=1.965(10)$ given by Berche \emph{et
al.}~\cite{berche-04} for the case $p=0.7$ mentioned above.

\begin{figure}[htbp]
\includegraphics*[width=9 cm]{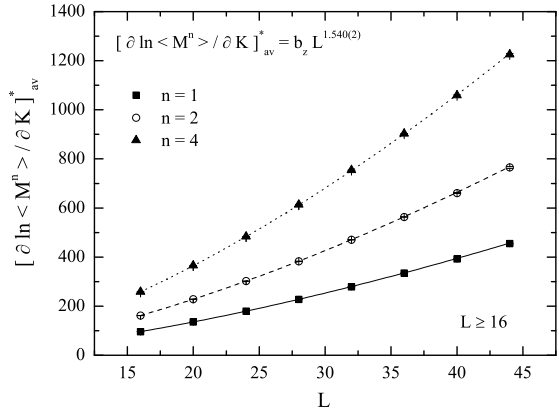}
\caption{\label{fig:3} FSS behavior of the peaks of the
logarithmic derivatives of the powers $n=1$, $2$, and $n=4$ of the
order parameter with respect to the inverse temperature. The
estimate for the exponent $1/\nu$ is given in the panel by
applying a simultaneous fitting attempt to a simple power law in
the size range $L=16-44$.}
\end{figure}

A similar simple power-law behavior was found for the peaks
corresponding to the absolute order-parameter derivative which
scale as $\sim L^{(1-\beta)/\nu}$. Since the examined behavior, in
this case, is very similar to the above behavior, we shall not
discuss further our efforts and give only our final estimate for
the relevant exponent $(1-\beta)/\nu=1.022(5)$. Combining the
above estimates and assuming at this point hyperscaling - in the
form: $2/\nu=2(1-\beta)/\nu+(d-\gamma/\nu)$ - we find
$1/\nu=1.54(1)$. In the following we will see that, this value is
not far from the effective estimates determined below from the
shift behavior of the system.

Let us now attempt the estimation of the correlation length
exponent via the scaling behavior of the logarithmic derivatives
of the powers $n=1$, $2$, and $n=4$ of the order parameter with
respect to the inverse temperature [see Eq.~(\ref{eq:4})]. Their
behavior was observed to be quite successfully fitted to a stable
simple power law with rather small variation of the effective
critical exponent. All the estimates were close to the above value
$1/\nu=1.54$ and for illustrative reasons we have presented one
such simultaneous fitting attempt in Fig.~\ref{fig:3}. Therefore,
the scheme appears to be in good agreement with hyperscaling. We
turn now to the estimation of this exponent from the general shift
behavior of the system.

As is well known, the pseudocritical temperatures, corresponding
to several finite-size anomalies, provide a traditional route for
the estimation of the critical temperature and the correlation
length exponent. A simultaneous fitting attempt to a power-law
shift behavior of the form $T_{[Z]_{av}^{\ast}}=T_{c}+b_{Z}\cdot
L^{-1/\nu}$ is often the attempted practice. For the present case
the following temperatures were calculated. Temperatures of the
peaks of the specific heat, magnetic susceptibility, inverse
temperature derivative of the absolute order parameter, and
inverse temperature logarithmic derivatives of the $n=1$, $2$, and
$n=4$ powers of the order parameter.
\begin{figure}[htbp]
\includegraphics*[width=9 cm]{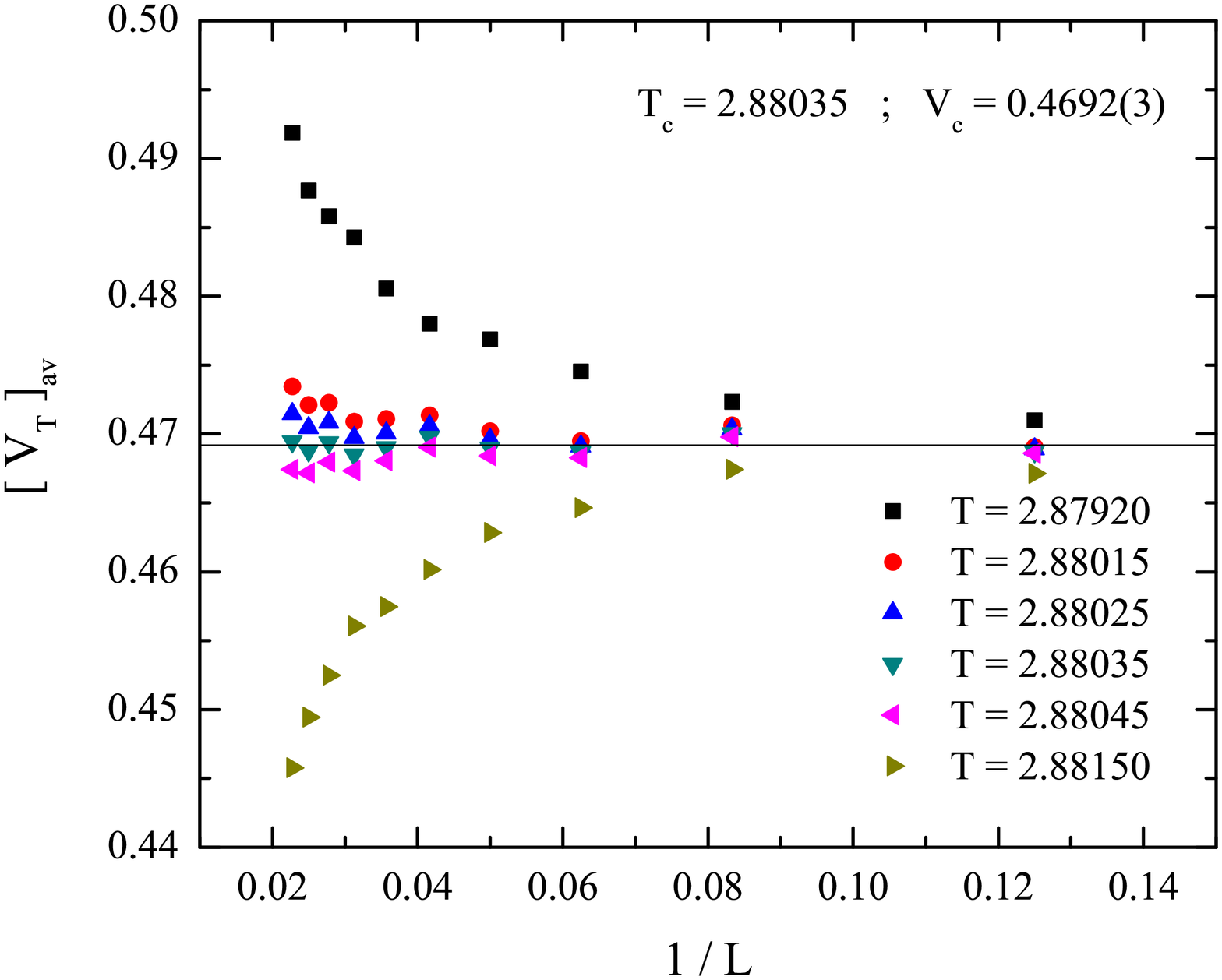}
\caption{\label{fig:4} (color online) Illustration of the
asymptotic trend of the fourth-order Binder's cumulant of the
order parameter for various temperatures close to the critical
temperature. Note the stability of the estimates at
$T_{c}=2.88035$.}
\end{figure}
The data were fitted in the ranges $L=8-44$ to $L=28-44$ and the
behavior of the estimates was observed. With the exception of the
fitting attempt corresponding to $L=8-44$, which gave an estimate
$1/\nu=1.465(1)$, all other attempts gave fluctuating estimates in
the range $1/\nu=1.52-1.54$, which seems to agree well with the
previous finding. Yet, at the same time, we observed a noticeable
shift of the estimated critical temperature from values
$T_{c}=2.8798$ to $T_{c}=2.88015$ as we varied $L_{min}$ from
$L_{min}=8$ to $L_{min}=28$. Here, a careful examination of the
behavior of the fourth-order Binder's cumulant of the order
parameter $[V_{T}]_{av}=[1-\langle M^4\rangle/3\langle
M^2\rangle^2]_{av}$ suggested that a larger critical temperature
in the range $T_{c}=2.88015 - 2.88045$ may be a very strong
option. Figure~\ref{fig:4} is a very clear illustration of this
statement, since it demonstrates that, at the temperature
$T_{c}=2.88035$, the behavior of the cumulant is almost
independent of $L$.

\begin{figure}[htbp]
\includegraphics*[width=9 cm]{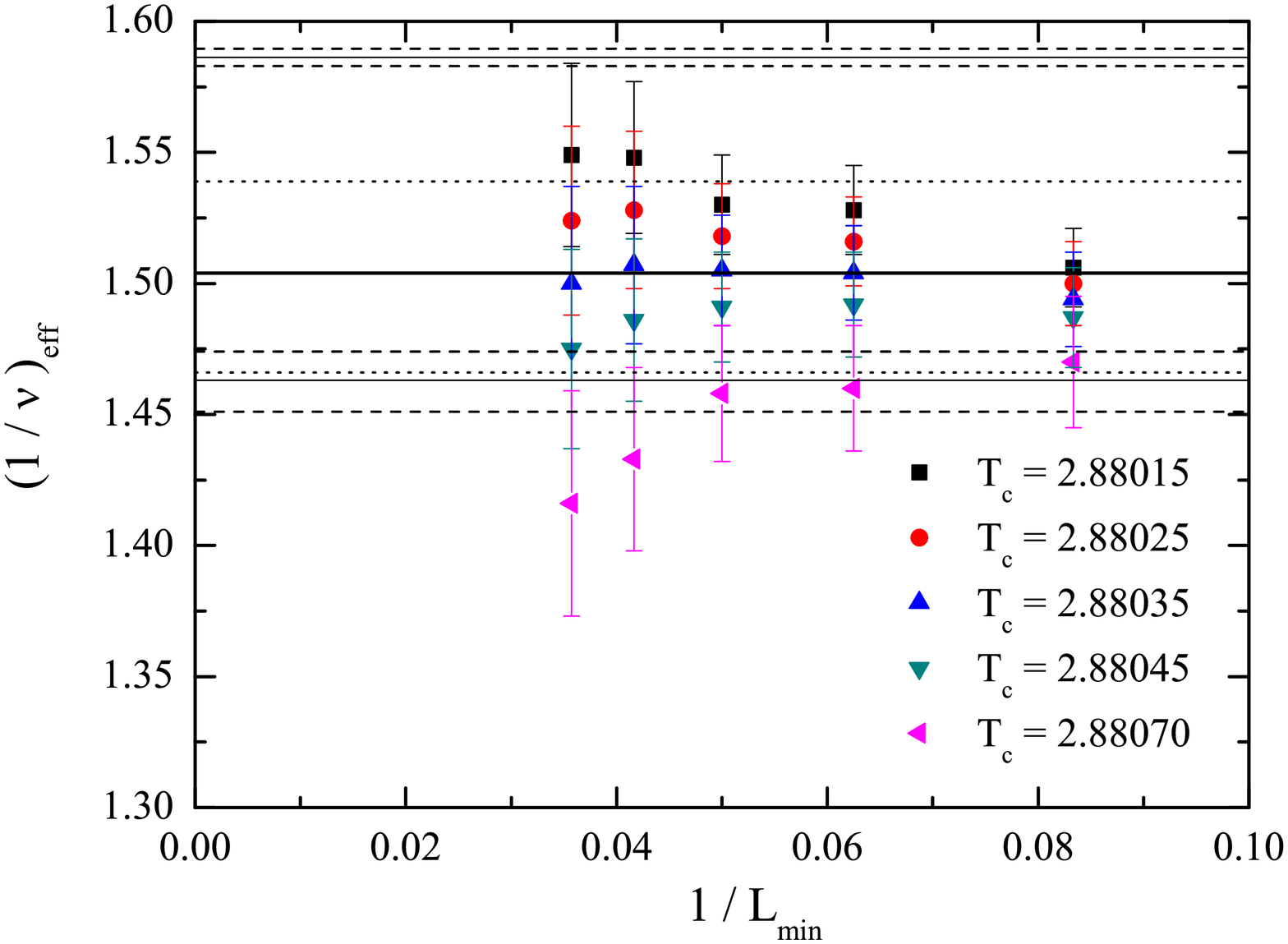}
\caption{\label{fig:5} (color online) A global illustration of the
estimates of the effective exponent $(1/\nu)_{eff}$. The solid
line drawn in the panel close to the value $1/\nu=1.586$, together
with the dashed lines, illustrate the location of the critical
exponent range for the pure 3d Ising model. The analogous range
for the 3d RIM is drawn close to the value $1/\nu=1.463$. The
range $1/\nu=1.466 - 1.539$ (dotted lines in the panel) around the
heavy solid line at the value $1/\nu=1.504$ demonstrate the
possible asymptotic evolution of the critical exponent for the
present model.}
\end{figure}

In order to better understand the shift-behavior of the system and
present a global illustration of it, we fix the critical
temperature to several values and calculate effective exponents.
We studied their size dependence by varying $L_{min}$ as usual,
from $L_{min}=12$ to $L_{min}=28$, and by further applying
simultaneous fitting attempts using the form
$T_{[Z]_{av}^{\ast}}=T_{c}+b_{Z}\cdot L^{-1/\nu}$ with fixed
critical temperatures from the set $T_c=\{2.88015,\;2.88025,\;
2.88035,\;2.88045,\;2.8807\}$. The resulting global behavior is
now illustrated in Fig.~\ref{fig:5}. This figure clearly indicates
that effective exponents are very sensitive to small temperature
changes. Furthermore, by applying a linear fitting in the
effective estimates for $L\geq 16$ for the case $T_{c}=2.88035$ we
obtain $1/\nu=1.504$ and this is indicated by the bold straight
line in the panel, as our central estimate for the exponent.
Repeating the same procedure for the case $T_{c}=2.88025$ we
obtain $1/\nu=1.539$, whereas for the case $T_{c}=2.88045$ we
obtain the value $1/\nu=1.466$. We may use these remote estimates
as upper and lower bounds respectively of our exponent estimation
and these are indicated in the panel by the doted lines. For the
sake of comparison we show also in the panel the accepted limits
for the estimation of the critical exponent $1/\nu$ of the pure 3d
Ising model $\nu=0.6304(13)$~\cite{Guida98} and the 3d RIM
$\nu=0.6837(53)$~\cite{ball-98}. Based on the above observations,
one could anticipate that any estimate in the range $1/\nu=1.466-
1.539$, indicated by the dotted lines in the panel, should be
acceptable. We point out here that, the exponent for the pure 3d
Ising model, $\nu=0.6304(13)$~\cite{Guida98} illustrated in
Fig.~\ref{fig:5}, is a moderate estimate with rather large error
bounds. A most recent and accurate estimate is $\nu=0.63002(10)$,
obtained by Hasenbusch~\cite{Hase10} in excellent agreement with
other recent studies of the 3d Ising model~\cite{Campo02,Bute02}.

Overall, the present case shares many of the problems encountered
by Berche \emph{et al.}~\cite{berche-04} in their study of the 3d
bond-diluted Ising model for the case of magnetic bond
concentration $p=0.7$. However, the above illustrations show that,
despite the already mentioned problems, our data are compatible
with the general expectations of the general 3d RIM. We close this
Section with two additional comments. Firstly, we also tried to
observe the shift behavior by fixing the shift exponent to the
value $1/\nu=1.463$ proposed by Ballesteros \emph{et
al.}~\cite{ball-98}.
\begin{figure}[htbp]
\includegraphics*[width=9 cm]{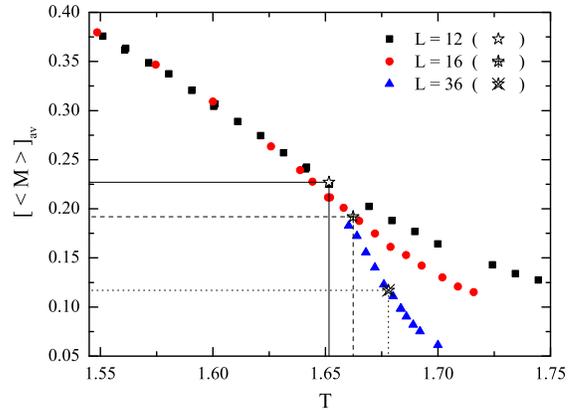}
\caption{\label{fig:6} (color online) Illustration of the
continuous behavior of the order parameter for the random-bond BC
model at $(\Delta,r)=(2.9,1/3)$. Three different lattice sizes
$L=12$, $16$, and $L=36$ are shown and the illustrated behavior is
approximately centered at the temperatures corresponding to the
peaks of the magnetic susceptibility, indicated by the asterisks
and the drop lines.}
\end{figure}
We then found that the corresponding simultaneous fitting attempts
gave values for the critical temperatures which are slowly
approaching the value $T_{c}=2.88035(10)$ suggesting once again
that this may be the asymptotic critical temperature. Secondly, we
observed the FSS behavior of the order parameter by fixing the
critical temperature to the values from the set
$T_{c}=\{2.88015,\;2.88025,\;2.88035,\; 2.88045\}$. The estimated
effective exponents $\beta/\nu$, had in all cases values in the
ranges $\beta/\nu=0.51(1)-0.52(1)$. These values, together with
the earlier estimate $\gamma/\nu=1.964(4)$, satisfy quite well
hyperscaling.

\section{Ex-first-order regime: Disorder-induced second-order phase transition}
\label{sec:4}

In this Section we investigate the effects of bond randomness on
the segment of the first-order transitions of the original pure
model. At the value $\Delta=2.9$, which is well inside the
first-order regime of the pure model, we have numerically verified
that the disorder strength $r=1/3$ is strong enough to convert,
without doubt, the original first-order transition to a genuine
second-order one. The illustration in Fig.~\ref{fig:6} consists
very clear evidence of this statement. In this figure we show the
behavior of the order parameter close to the pseudocritical region
for three different lattice sizes $L=12$, $16$, and $L=36$. The
three different asterisks and the respective drop lines indicate
the points on these curves corresponding to the magnetic
susceptibility peaks. These order-parameter curves, at the
corresponding pseudocritical regions, are smooth and no sign of
discontinuity appears as the lattice size increases to $L=36$.
Since the order parameter is continuous at the combination
$(\Delta,r)=(2.9,1/3)$, the disorder-induced transition is a
second-order phase transition, which we will take as
representative of the ex-first-order universality class.

For the case $(\Delta,r)=(2.9,1/3)$, we start the FSS analysis by
attempting the estimation of the exponent ratio that characterizes
the divergence of the susceptibility. First, we assume that the
(sample-averaged) susceptibility obeys a simple power law of the
form $[\chi_T]_{av}^{\ast}=b\cdot L^{\gamma/\nu}$. However, the
fitting attempts are found to be unstable with respect to the size
range (we attempted the ranges $L=8-44,\; 12-44,\;\ldots$) and the
resulting coefficients $b$ and exponents $\gamma/\nu$ are varying
in a competitive way, indicating the need of correction terms.
\begin{figure}[htbp]
\includegraphics*[width=9 cm]{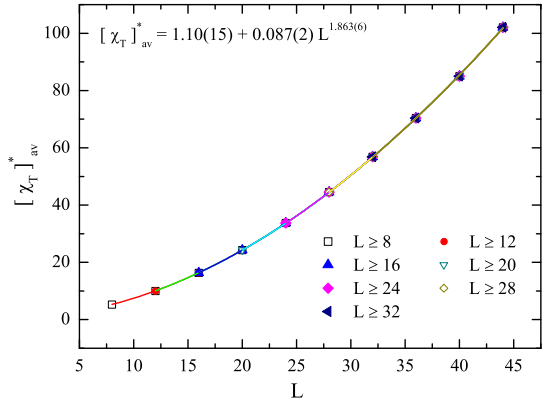}
\caption{\label{fig:7} (color online) The divergence of the
susceptibility maxima is well described by a scaling law of the
form $[\chi_{T}]_{av}^{\ast}=a+b\cdot L^{\gamma/\nu}$.
Illustration of a simultaneous global fitting attempt on the data
corresponding to several size ranges shown in the panel, indicated
by the different line colors. Note that, the simultaneous attempt
gives an estimate of $a=1.10(15)$ which is in agreement, within
errors, with all separate attempts and effectively coincides with
the estimates of the ranges $L=8-44$ and $L=12-44$.}
\end{figure}
Therefore, we tried to find a resolution of this problem by
introducing correction terms and we found that a constant
background term, was actually the most effective in eliminating
the observed instability. The behavior of the corresponding
fitting attempts, assuming the scaling relation
$[\chi_T]_{av}^{\ast}=a+b\cdot L^{\gamma/\nu}$, is very good and
produces very slowly varying fitting parameters. A completely
stable behavior was obtained by fixing the value of the background
term to be of the order of $a=1.0-1.2$. The values of the
background term for the first two ranges $L=8-44$ and $L=12-44$,
and the value obtained by a simultaneous global fitting attempt on
the data corresponding to several size ranges, as shown in
Fig.~\ref{fig:7}, are effectively the same. As indicated in the
panel of this figure the global attempt gives an estimate
$a=1.10(15)$. In the same panel we give the estimates of
$b=0.087(2)$ and the exponent $\gamma/\nu=1.863(6)$. Using this
value ($a=1.1$) we illustrate in Fig.~\ref{fig:8} the stability of
the resulting scaling scheme by presenting, in a double
logarithmic scale, the behavior of the two most remote ranges
$L=8-44$ and $L=28-44$. The coincidence of the estimates in this
panel appears as a guaranty of the quite accurate estimation of
the magnetic exponent ratio $\gamma/\nu=1.864(12)$.

However, it may be crucial for future studies to give here a more
detailed discussion on the various corrections tested before
adopting the above scenario.
\begin{figure}[htbp]
\includegraphics*[width=9 cm]{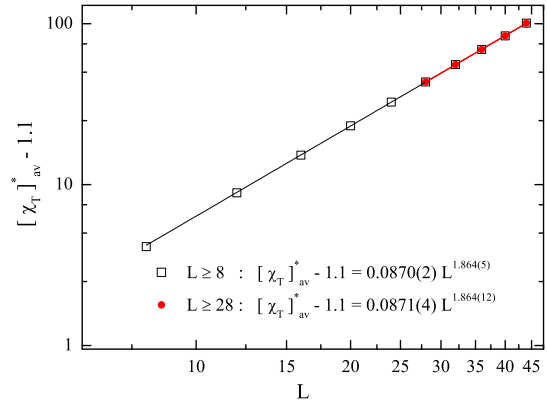}
\caption{\label{fig:8} (color online) This figure is complementary
to Fig.~\ref{fig:7} and further illustrates the stability of the
scaling scheme by presenting, in a log-log scale, the behavior of
the two most remote ranges $L=8-44$ and $L=28-44$. Note that, we
have here subtracted from the susceptibility data the estimated
value of the background term ($a=1.1$).}
\end{figure}
Our fitting attempts followed a quite common practice restricting
the search to an expression including only one correction term,
e.g., $[\chi_{T}]_{av}^{\ast}=b\cdot L^{\gamma/\nu}+b'\cdot
L^{\gamma/\nu-\epsilon}$. This restriction was unavoidable, since
the fits were already notoriously unstable, as we increased the
minimum lattice size, with the above four-parameter expression.
The stability and quality of the fittings were then observed by
fixing the correction exponent $\epsilon$ to various relevant
values and treating only the exponent $\gamma/\nu$ as a free one.
The quality of the fittings was characterized by their cumulated
square deviation, $\chi^{2}$, and the stability of the values of
the exponent $\gamma/\nu$ and the corresponding amplitudes were
observed. The best sequence of fittings was obtained for
$\epsilon=\gamma/\nu$, corresponding to the regular background
behavior adopted in this paper and illustrated in
Figs.~\ref{fig:7} and \ref{fig:8}. This choice is unique, in the
sense that is for sure the simplest one. Besides, all the other
tested values of $\epsilon$, in the neighborhood of the leading
and next-to-leading corrections to scaling, corresponding to the
3d RIM $\epsilon=\omega=0.33(3)$, $\epsilon=2\omega=0.66(6)$ and
$\epsilon=\omega_{2}=0.82(8)$~\cite{HaseP07}, did not produce a
stable sequence of fittings, but gave a rather pathological
variation of the estimates of the exponent $\gamma/\nu$ and the
relevant amplitudes. Yet, one may also observe the stability and
quality of the fittings by fixing both the exponent $\gamma/\nu$
and the exponent $\epsilon$. In these final attempts, we did found
cases of comparable quality and stability that are completely
compatible with the 3d RIM universality class. One such case is
described by the expression $[\chi_{T}]_{av}^{\ast}=0.0494(4)\cdot
L^{1.964}+0.013(1)\cdot L^{1.964-0.66}$, and another one is
described by the expression $[\chi_{T}]_{av}^{\ast}=0.052(2)\cdot
L^{1.964}+0.18(2)\cdot L^{1.964-0.82}$. The above two expressions
and the behavior in Fig.~\ref{fig:8} produce very close values
(almost identical within statistical errors) in the range $L=8-44$
studied in this paper. This observation should serve also as a
warning of the difficulties and the pathology of the fitting
attempts, since the existence of stable forms, with quite
different correction exponents, means also that a completely
reliable estimation of corrections-to-scaling exponents may not
been feasible even at larger lattice sizes.
\begin{figure}[htbp]
\includegraphics*[width=9 cm]{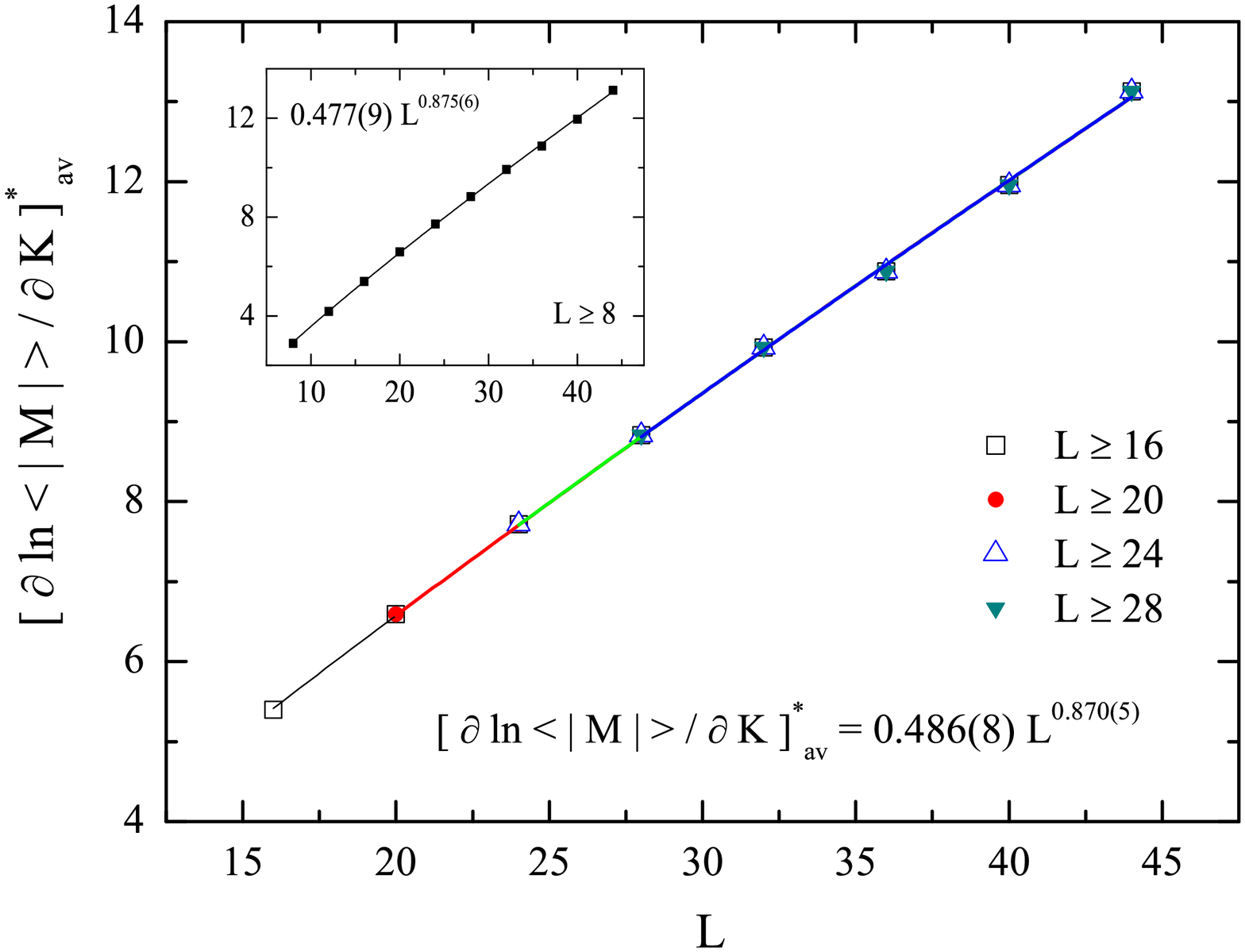}
\caption{\label{fig:9} (color online) Illustration of a
simultaneous global fitting attempt on the data corresponding to
several size ranges shown in the panel for the maxima of the
sample-averaged absolute order-parameter derivative. Note that in
this case, the fitting attempts are insensitive to the size ranges
used.}
\end{figure}

Let us now investigate the FSS behavior of the peaks corresponding
to the absolute order-parameter derivative which, as mentioned
earlier, are expected to scale as $\sim L^{(1-\beta)/\nu}$ with
the lattice size. Here, we find that, the corresponding maxima
obey very well a simple power law $[\partial \langle |M|\rangle /
\partial K]_{av}^{\ast}=b\cdot L^{(1-\beta)/\nu}$ without any
correction terms. The corresponding fitting attempts are very
stable with respect to the lattice-size range, as moving from
$L=8-44$ to $L=28-44$. Figure~\ref{fig:9} illustrates in the main
panel a simultaneous global fitting attempt on the data
corresponding to the size ranges shown. The inset of this figure
presents a simple fitting for the complete lattice range $L=8-44$.
The estimates for the exponent are almost insensitive to the used
lattice-size ranges giving for the simultaneous global fitting
$(1-\beta)/\nu=0.870(5)$ and for the simple fitting $L=8-44$, in
the inset, $(1-\beta)/\nu=0.875(6)$. These results indicate the
consistency of the FSS scheme and the accuracy of the numerical
data. We may take as a final (quite confident and moderate in its
error bounds) estimate, the value $(1-\beta)/\nu=0.87(1)$.

\begin{figure}[htbp]
\includegraphics*[width=9 cm]{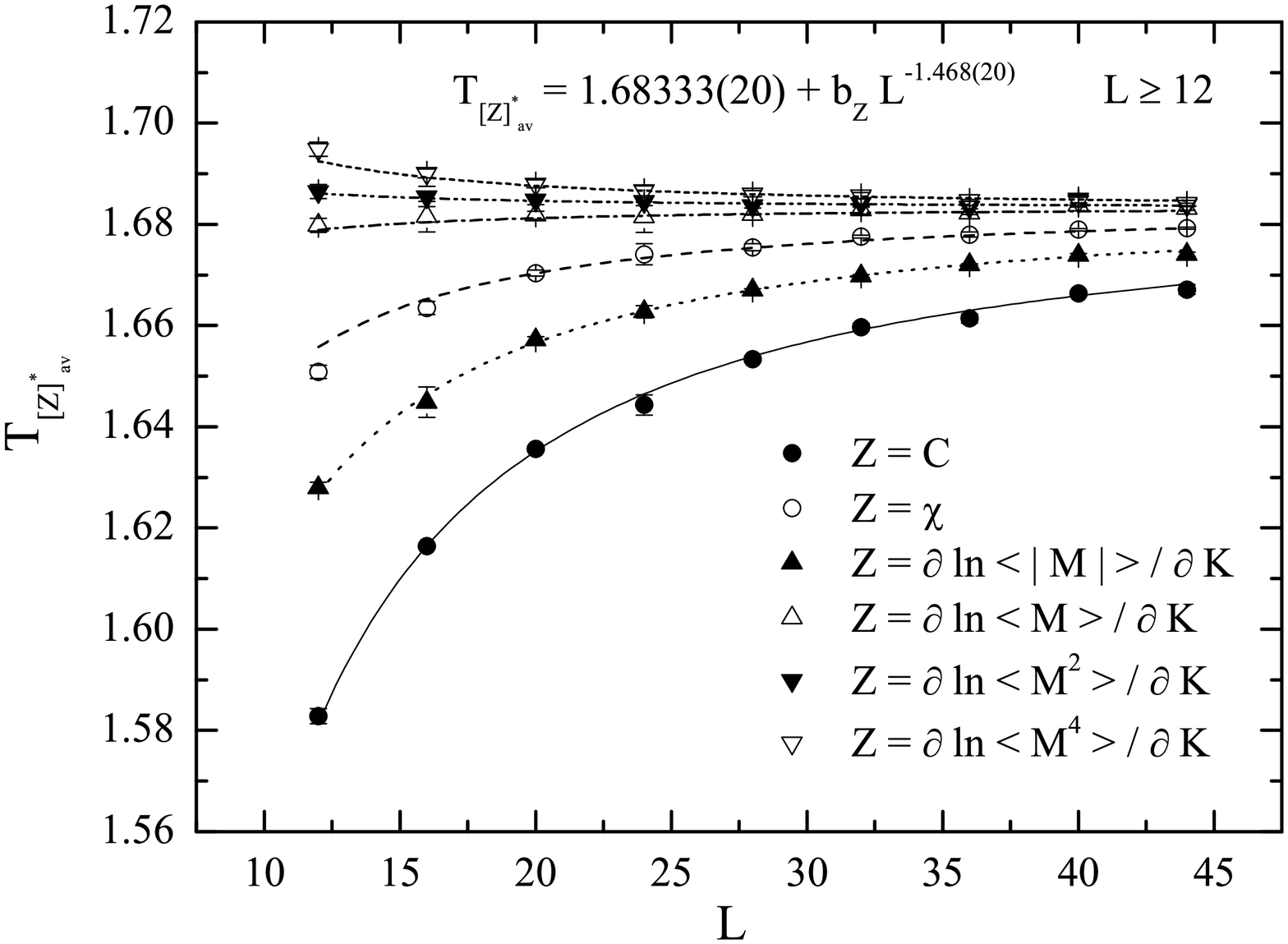}
\caption{\label{fig:10} FSS behavior of the pseudocritical
temperatures defined in the text. Estimates for the critical
temperatures and the shift exponent $1/\nu$ are given in the
panel. The stability of the fitting scheme is discussed in the
relevant text.}
\end{figure}

As mentioned earlier, the pseudocritical temperatures,
corresponding to several finite-size anomalies, provide a route
for the estimation of the critical temperature and the correlation
length exponent. A simultaneous fitting attempt to a power-law
shift behavior of the form $T_{[Z]_{av}^{\ast}}=T_{c}+b_{Z}\cdot
L^{-1/\nu}$ is the generally suggested practice.
Figure~\ref{fig:10} illustrates the shift behavior of such several
pseudocritical temperatures.
These temperatures correspond to the peaks of the following six
(sample-averaged) quantities: specific heat, magnetic
susceptibility, inverse temperature derivative of the absolute
order parameter, and inverse temperature logarithmic derivatives
of the $n=1$, $n=2$, and $n=4$ powers of the order parameter. The
data illustrated are fitted in the range $L=12-44$, which
corresponds to the best fitting attempt of all tried. The
resulting estimates of the critical temperatures and the shift
exponent $1/\nu$ are given in the panel. However, in this case
also, the fitting attempts for the estimation of the critical
exponent $1/\nu$ from the data of the pseudocritical temperatures
are not completely stable. Therefore, some further comments and
analysis are necessary. Despite that, the estimation of the
critical temperature is quite stable, with values ranging from
$T_{c}=1.6833(2)$ to $T_{c}=1.6835(2)$, as we vary the
lattice-size ranges from $L=8-44$ to $L=24-44$. However, the
corresponding exponent estimates $1/\nu$ have a noticeable
variation with estimates from $1.51(2)$ to $1.42(2)$. The
statistical errors influence here the quality but also the
stability of the fitting attempts.

To better understand the shift-behavior we tried the following two
assumptions. First, we fixed the critical temperature to be
$T_{c}=1.6835$, a value indicated also by a very careful
examination of the behavior of the fourth-order Binder's cumulant
of the order parameter (not shown here for brevity). Again, we
found a similar variation as above with the lattice-size ranges
used. The best fittings were obtained for the ranges $L=12-44$ and
$L=16-44$, giving also very close estimates for the shift exponent
$1/\nu$ [$1/\nu=1.45(2)$]. Subsequently, we tried to observe the
behavior of the estimates of the critical temperature by fixing
the shift exponent to the value $1/\nu=1.438$. This is the value
obtained by satisfying hyperscaling, given the previous estimates
for $\gamma/\nu=1.864$ and $(1-\beta)/\nu=0.87$.
\begin{figure}[htbp]
\includegraphics*[width=9 cm]{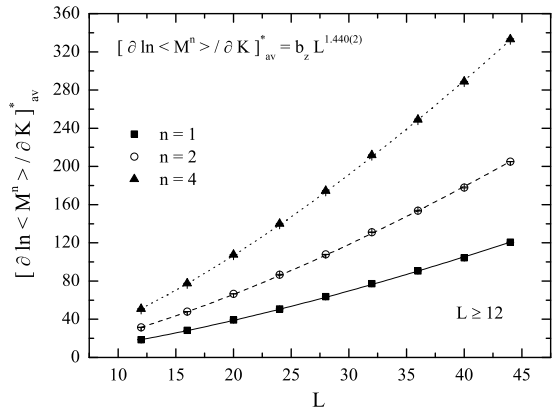}
\caption{\label{fig:11} FSS behavior of the peaks of the
logarithmic derivatives of the powers $n=1$, $2$, and $n=4$ of the
order parameter with respect to the inverse temperature. The
estimate for the exponent $1/\nu$ is given in the panel by
applying a simultaneous fitting attempt to a simple power law in
the size range $L=12-44$.}
\end{figure}
The estimates for the critical temperature are all very close to
$T_{c}=1.68345$ and the best fitting attempt, giving also an
estimate with the smallest error, is $T_{c}=1.6835$, corresponding
to the range $L=12-44$. We have therefore accepted as our best
estimation of the shift behavior the values $T_{c}=1.6835(2)$ and
$1/\nu=1.45(2)$.

An independent estimation of the correlation length exponent can
be obtained via the scaling behavior of the logarithmic
derivatives of the powers $n=1$, $2$, and $n=4$ of the order
parameter with respect to the inverse temperature [see
Eq.~(\ref{eq:4})]. Their behavior was observed to be quite stable
and consistent with the above estimation from the shift behavior.
We tried here to vary again the ranges from $L=8-44$ to $L=28-44$.
Depending on $L_{min}$ the estimated effective exponents vary in
the range $1/\nu = 1.445 - 1.415$. In Fig.~\ref{fig:11} we
illustrate such an estimation by a simultaneous fitting attempt in
the range $L=12-44$. As it can be seen from this figure the
estimate is $1/\nu=1.440(2)$. Therefore, from this FSS scheme one
should conclude that $1/\nu=1.430(10)$ and combining with the
above shift behavior, we should regard $1/\nu=1.440(10)$ a very
decent proposal that is now in full agreement with the exponent
value $1/\nu=1.438$ obtained by satisfying hyperscaling, given the
previous estimates for $\gamma/\nu=1.864$ and
$(1-\beta)/\nu=0.87$.

\begin{figure}[htbp]
\includegraphics*[width=9 cm]{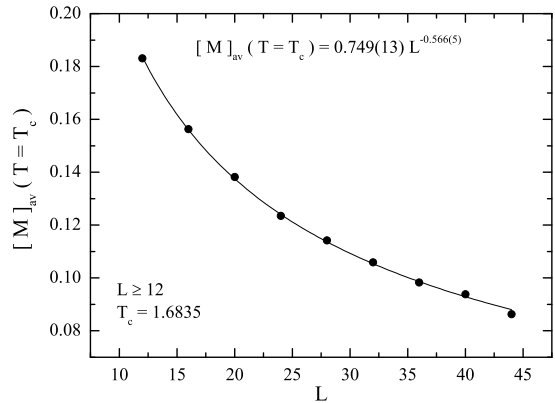}
\caption{\label{fig:12} FSS behavior of the order parameter at the
estimated critical temperature. In the panel we show a simple
power-law estimation of the exponent ratio $\beta/\nu$.}
\end{figure}

Finally we give an outline on the behavior of the order parameter
at the estimated critical temperature. We computed the finite-size
values of the order parameter at the temperature $T_{c}=1.6835$
from the corresponding (sample-averaged) order-parameter curves.
In Fig.~\ref{fig:12} we apply a simple power-law estimation for
the exponent ratio $\beta/\nu$, using again the size range
$L=12-44$. This estimation gives a critical exponent ratio
$\beta/\nu=0.566(5)$ in excellent agreement with the value
$\beta/\nu=0.568$, which is obtained by satisfying hyperscaling,
given the estimates $\gamma/\nu=1.864$ and $(1-\beta)/\nu=0.87$.
Furthermore, it should be noted that the total variation with the
lattice-size range is very small, ranging from $\beta/\nu=0.56(1)$
to $\beta/\nu=0.57(2)$, as we vary the size range from $L=8-44$ to
$L=24-44$.

Summarizing, our findings in this Section on the emerging, under
bond randomness, second-order phase transition of the 3d
random-bond BC model are the following: (i) the proposed critical
exponents provide a stable finite-size behavior, strongly
supporting hyperscaling and (ii) the proposed value of the
critical exponent $\gamma/\nu=1.864(12)$ characterizes, in a very
clear way, the expected distinctive strong-coupling fixed point,
describing, according to the renormalization-group calculations,
the emerging from the first-order regime second-order phase
transition.

\section{Summary and Conclusions}
\label{sec:5}

\begin{table*}
\caption{\label{tab:1}Summary of critical exponents for the 3d
pure and disordered Ising (IM), $q$-states Potts (PM), and
Blume-Capel (BCM) models, as obtained in
Refs.~\cite{Guida98,Hase10,ball-98,HaseP07,berche-04,ball-00,chate-01}
and the present paper.}
\begin{ruledtabular}
\begin{tabular}{lcccccc}
Model &$1/\nu$ &$\nu$ &$\gamma/\nu$ &$\eta=2-\gamma/\nu$ &$\beta/\nu$ &$(2\beta/\nu)+\gamma/\nu$\\
\hline
 Ex-second-order phase transition\\
 \hline
 Pure IM~\cite{Guida98}                             & 1.586(3)  & 0.6304(13) & 1.966(3)  & 0.034(3)  & 0.517(3)  & 3.00(9)   \\
 Pure IM~\cite{Hase10}                                   &      & 0.63002(10)&           & 0.03627(10) &           &         \\
 Site-diluted IM($p=0.8$)\footnotemark[1]~\cite{ball-98}                       & 1.463(11) & 0.6837(53) & 1.963(5)  & 0.037(5)  &           &          \\
 Site-diluted IM($p=0.8$)\footnotemark[1]~\cite{HaseP07}                           &       & 0.683(2)   &           & 0.036(1)  &           &          \\
 Bond-diluted IM ($p=0.7$)\footnotemark[1]~\cite{berche-04}        & 1.520(20)   & 0.660(10)    & 1.965(10) & 0.035(10) & 0.515(5)  & 2.995(20) \\
 Bond-diluted IM ($p=0.55$)\footnotemark[1]~\cite{berche-04}       & 1.460(20)   & 0.685(10)  & 1.977(10) & 0.023(10) & 0.513(5)  & 3.003(20) \\
 Random-bond BCM ($\Delta=1$; $r=1/3$)               & 1.504(19) & 0.665(17)  & 1.964(4)  & 0.036(4)  & 0.510(10) & 2.984(24) \\
 \hline
Ex-first-order phase transition\\
 \hline
 Bond-diluted $q=3$ PM~\cite{ball-00}               & 1.449(10) & 0.690(5)   & 1.922(4)  & 0.078(4)  &           &           \\
 Random-bond BCM ($\Delta=2.9$; $r=1/3$)            & 1.440(10) & 0.694(5)   & 1.864(12) & 0.136(12) & 0.566(5)  & 2.996(22)\\
 Bond-diluted $q=4$ PM~\cite{chate-01}          & 1.330(25) & 0.752(14)  & 1.500(14) & 0.500(14) & 0.645(24) & 2.790(62) \\
\end{tabular}
\end{ruledtabular}
 \footnotetext[1]{$1-p$ denotes the concentration of impurities.}
\end{table*}

It is instructive at this point to attempt an overview on the
effects of disorder for 3d ex-second- and ex-first-order phase
transitions and compare our results with previous relevant
research. Let us restrict ourselves to a presentation focused
mainly on some of the papers discussed already in the text. These
are the cases of the 3d RIM~\cite{ball-98,berche-04,HaseP07} and
those of the ex-weak first-order phase transition of the 3d
site-diluted $q=3$ Potts model studied by Ballesteros \emph{et
al.}~\cite{ball-00}, and the case of the ex-strong first-order
transition of the 3d bond-diluted $q=4$ Potts model studied by
Chatelain \emph{et al.}~\cite{chate-01}. In table~\ref{tab:1} we
display the critical exponents obtained in these papers together
with the two concrete cases of the random-bond 3d BC Model. From
this table one can see that, for the two cases studied here, the
hyperscaling relation $(2\beta/\nu)+\gamma/\nu=d$ is well
satisfied, and the best case is that of the ex-first-order
transition that was also found to have a very robust FSS behavior.
Furthermore, a straightforward comparison shows that, our
ex-second-order case ($\Delta=1$; $r=1/3$) has a very similar
behavior with that of Berche \emph{et al.}~\cite{berche-04} on the
3d bond-diluted Ising model with magnetic bond concentration
$p=0.7$. This is quite remarkable and, we may also point out that,
our results are limited to lattice sizes up to $L=44$, whereas
sizes up to $L=96$ have been simulated in that paper. Our efforts
indicate, in agreement with Berche \emph{et al.}~\cite{berche-04},
that at these lattice sizes the system still crosses over to the
universality class of the RIM, described by the second entry in
table~\ref{tab:1}~\cite{ball-98}. Finally, we have deliberately
placed our second case of study of the random-bond BC Model
($\Delta=2.9$; $r=1/3$) after the ex-weak- and before the
ex-strong-first-order transitions of the 3d Potts model, since our
results for the critical exponents appear to interpolate between
these two cases.

In conclusion, the 3d random-bond BC model has been studied
numerically in both its first- and second-order phase transition
regimes by a comprehensive FSS analysis. As expected on general
universality arguments the 3d random-bond BC model at the
second-order regime ($\Delta=1$) was found to be fully compatible
with the 3d Ising universality class. However, the case studied
here $[(\Delta,r)=(1,1/3)]$ exhibits analogous crossover problems
as the ones encountered in the case of the bond-diluted 3d Ising
model with magnetic bond concentration $p=0.7$ studied by Berche
\emph{et al.}~\cite{berche-04}. For the case of ex-first-order
regime at $\Delta=2.9$, we have shown that the disorder strength
$r=1/3$ was strong enough to convert, without doubt, the original
first-order transition to a genuine second-order one. For this
case $[(\Delta,r)=(2.9,1/3)]$ we have presented a detailed and
convincing FSS scheme. The scenario adopted in this paper, and the
proposed critical exponents obtained from a stable scaling
behavior, are supporting strongly hyperscaling. In particular the
value of the critical exponent $\gamma/\nu=1.864(12)$ is robust
and characterizes, in a very clear way, the expected distinctive
strong-coupling fixed point, describing, according to the
renormalization-group calculations, the emerging from the
first-order regime second-order pase transition. The present
results point out, as in the case of the 2d random-bond BC model,
the existence of a strong violation of universality principle of
critical phenomena, since the two second-order transitions between
the same ferromagnetic and paramagnetic phases have different sets
of critical exponents. In the 2d random-bond BC model the
difference in the exponents was revealed in the thermal exponent
$\nu$ and an extensive but weak universality was found
corresponding to the fact that the two emerging transitions have
the same magnetic exponent ratios ($\beta/\nu$ and
$\gamma/\nu$)~\cite{malakis09}. However, for the 3d random-bond BC
model no such weak universality is supported. The observed strong
violation of universality is now revealed mainly, but not
exclusively, in the fact that the corresponding emerging
transitions have different magnetic exponent ratios, as seen from
the values of $\gamma/\nu$ of table~\ref{tab:1}.

The proposal for the possible new universality class in the 3d
random BC model in the ex-first-order regime is an interesting
finding supported also by the early renormalization-group
calculations~\cite{hui89,falicov96}, as mentioned already in the
introduction. It is also a surprising result since the two
transitions are between the same ferromagnetic and paramagnetic
phases. However, having in mind that the 3d RIM suffers rather
slowly decaying scaling
corrections~\cite{ball-98,berche-04,HaseP07}, one can never be
confident for the asymptotic behavior at the present system sizes.
In view of the above comments, and the earlier discussion of our
fitting attempts for the susceptibility maxima, a confident
resolution of this situation may require lattice sizes of the
order of at least $L=240$. Further investigations, for instance
considering the behavior for different values of the crystal-field
coupling $\Delta$, suitably chosen in the ex-first-order regime,
may be very useful to this direction. In a more advanced level one
may try to obtain further convincing evidence by comparing
multifractal spectra for correlations
functions~\cite{Ludw90,Month09} for the ex-second- and
ex-first-order transitions. These lines of research demand further
computational efforts and more sophisticated MC and FSS schemes,
but as pointed out in subsection~\ref{sec:2b}, may yield more
precise and additional useful information concerning the
properties of disorder averages~\cite{wiseman-98}.

\begin{acknowledgments}
The authors are grateful to V. Mart\'{i}n-Mayor for a critical
reading of the manuscript. This work was supported by the special
Account for Research of the University of Athens (code: 11112).
N.G. Fytas has been partly supported by MICINN, Spain, through a
research contract No. FIS2009-12648-C03.
\end{acknowledgments}

{}


\begin{thebibliography}{}
\bibitem{harris74} A.B. Harris, J. Phys. C {\bf 7}, 1671 (1974).
\bibitem{berker90} A.N. Berker, Phys. Rev. B {\bf 42}, 8640 (1990).
\bibitem{aizenman89} M. Aizenman and J. Wehr, Phys. Rev. Lett. {\bf 62}, 2503 (1989); {\bf 64}, 1311(E) (1990).
\bibitem{hui89} K. Hui and A.N. Berker, Phys. Rev. Lett. {\bf 62}, 2507 (1989); {\bf 63}, 2433(E) (1989).
\bibitem{berker93} A.N. Berker, Physica A {\bf 194}, 72 (1993).
\bibitem{Cardy97} J. Cardy and J.L. Jacobsen, Phys. Rev. Lett. {\bf 79}, 4063 (1997); J.L. Jacobsen and J. Cardy, Nucl. Phys. B {\bf 515}, 701 (1998);
J. Cardy, Physica A {\bf 263}, 215 (1998).
\bibitem{Ferna12} L.A. Fern\'{a}ndez, A. Gordillo-Guerrero, V.Mart\'{i}n-Mayor, and J.J. Ruiz-Lorenzo, arXiv:1205.0247.
\bibitem{greenblatt09} R.L. Greenblatt, M. Aizenman and J.L. Lebowitz, Phys. Rev. Lett. {\bf 103}, 197201 (2009);
Physica A {\bf 389}, 2902 (2010).
\bibitem{chen92} S. Chen, A.M. Ferrenberg, and D.P. Landau, Phys. Rev. Lett. {\bf 69}, 1213 (1992).
\bibitem{falicov96} A. Falicov and A.N. Berker, Phys. Rev. Lett. {\bf 76}, 4380 (1996).
\bibitem{malakis09} A. Malakis, A.N. Berker, I.A. Hadjiagapiou, and N.G. Fytas, Phys. Rev. E {\bf 79}, 011125
(2009); A. Malakis, A.N. Berker, I.A. Hadjiagapiou, N.G. Fytas,
and T. Papakonstantinou, \emph{ibid.} {\bf 81}, 041113 (2010).
\bibitem{wang-01} F. Wang and D.P. Landau, Phys. Rev. Lett. {\bf 86}, 2050
(2001); Phys. Rev. E {\bf 64}, 056101 (2001).
\bibitem{landau-80} D.P. Landau, Phys. Rev. B {\bf 22}, 2450 (1980).
\bibitem{chow-86} D. Chowdhury and D. Stauffer, J. Stat. Phys. {\bf 44}, 203 (1986).
\bibitem{heuer-90} H.-O. Heuer, Europhys. Lett. {\bf 12}, 551 (1990); Phys. Rev. B {\bf 42}, 6476
(1990); J. Phys. A {\bf 26}, L333 (1993).
\bibitem{hennecke-93} M. Hennecke, Phys. Rev. B {\bf 48}, 6271 (1993).
\bibitem{ball-98} H.G. Ballesteros, L.A. Fern\'{a}ndez, V. Mart\'{i}n-Mayor, A. Mu\~{n}oz Sudupe, G. Parisi,
and J.J. Ruiz-Lorenzo, Phys. Rev. B {\bf 58}, 2740 (1998).
\bibitem{wiseman-98} S. Wiseman and E. Domany, Phys. Rev. Lett. {\bf 81}, 22 (1998); Phys. Rev. E {\bf 58}, 2938 (1998).
\bibitem{calabrese-03} P. Calabrese, V. Mart\'{i}n-Mayor, A. Pelissetto, and E. Vicari,
Phys. Rev. E {\bf 68}, 036136 (2003).
\bibitem{HaseP07} M. Hasenbusch, F. Parisen Toldin, A. Pelissetto, and E. Vicari, J. Stat. Mech.: Theory Exp.
(2007) P02016.
\bibitem{berche-02} P.E. Berche, C. Chatelain, B. Berche, and W. Janke, Comp. Phys. Comm. {\bf 147}, 427 (2002).
\bibitem{berche-04} P.E. Berche, C. Chatelain, B. Berche, and W. Janke, Eur. Phys. J. B {\bf 38}, 463 (2004).
\bibitem{Iva-05} D. Ivaneyko, J. IInytskyi, B. Berche, and Yu. Holovatch, Cond. Matt. Phys. {\bf 8}, 149 (2005).
\bibitem{fytas-10} N.G. Fytas and P.E. Theodorakis, Phys. Rev. E {\bf 82}, 062101 (2010).
\bibitem{Hase07} M. Hasenbusch, F. Parisen Toldin, A. Pelissetto, and E. Vicari, Phys. Rev. B {\bf 76}, 094402 (2007).
\bibitem{folk-00} R. Folk, Y. Holovatch, and T. Yavors'kii, Phys. Rev. B {\bf 61}, 15114 (2000).
\bibitem{pakhnin-00} D.V. Pakhnin and A.I. Sokolov, Phys. Rev. B {\bf 61}, 15130 (2000).
\bibitem{pelissetto-00} A. Pelissetto and E. Vicari, Phys. Rev. B {\bf 62}, 6393 (2000).
\bibitem{newman-82} K.E. Newman and E.K. Riedel, Phys. Rev. B {\bf 25}, 264 (1982).
\bibitem{jug-83} G. Jug, Phys. Rev. B {\bf 27}, 609 (1983).
\bibitem{mayer-89} I.O. Mayer, J. Phys. A {\bf 22}, 2815 (1989).
\bibitem{ball-00} H.G. Ballesteros, L.A. Fern\'{a}ndez, V. Mart\'{i}n-Mayor, A. Mu\~{n}oz Sudupe, G. Parisi,
and J.J. Ruiz-Lorenzo, Phys. Rev. B 61, 3215 (2000).
\bibitem{chate-01} C. Chatelain, B. Berche, W. Janke,and P.E. Berche, Phys. Rev. E {\bf 64}, 036120 (2001); Nucl. Phys. B {\bf 719}, 725 (2005).
\bibitem{Puha-01} I. Puha and H.T. Diep, J. Magn. Magn. Mater. {\bf 224}, 85 (2001).
\bibitem{Salmon-10} O.D.R. Salmon and J.R Tapia, J. Phys. A {\bf 43}, 125003 (2010).
\bibitem{Wu-10} X.T. Wu, Phys. Rev. E {\bf 82}, 010101 (2010).
\bibitem{Dias-11} D.A. Dias and J.A. Plascak, Phys. Lett. A {\bf 375}, 2089 (2011).
\bibitem{blume66} M. Blume, Phys. Rev. {\bf 141}, 517 (1966).
\bibitem{capel66} H.W. Capel, Physica (Utr.) {\bf 32}, 966 (1966); {\bf 33}, 295 (1967); {\bf 37}, 423 (1967).
\bibitem{landau72} D.P. Landau, Phys. Rev. Lett. {\bf 28}, 449
(1972); A.N. Berker and M. Wortis, Phys. Rev. B {\bf 14}, 4946
(1976); M. Kaufman, R.B. Griffiths, J.M. Yeomans and M. Fisher,
\emph{ibid}. {\bf 23}, 3448 (1981); W. Selke and J. Yeomans, J.
Phys. A {\bf 16}, 2789 (1983); D.P. Landau and R.H. Swendsen,
Phys. Rev. B {\bf 33}, 7700 (1986); J.C. Xavier, F.C. Alcaraz, D.
Pena Lara, and J.A. Plascak, \emph{ibid.} {\bf 57}, 11575 (1998).
\bibitem{stephen73} M.J. Stephen and J.L. McColey, Phys. Rev.
Lett. {\bf 44}, 89 (1973); T.S. Chang, G.F. Tuthill, and H.E.
Stanley, Phys. Rev. B {\bf 9}, 4482 (1974); G.F. Tuthill, J.F.
Nicoll, and H.E. Stanley, \emph{ibid.} {\bf 11}, 4579 (1975); F.J.
Wegner, Phys. Lett. {\bf 54A}, 1 (1975).
\bibitem{fox73} P.F. Fox and A.J. Guttmann, J. Phys. C {\bf 6},
913 (1973); T.W. Burkhardt and R.H. Swendsen, Phys. Rev. B {\bf
13}, 3071 (1976); W.J. Camp and J.P. Van Dyke, \emph{ibid.} {\bf
11}, 2579 (1975); D.M. Saul, M. Wortis, and D. Stauffer,
\emph{ibid.} {\bf 9}, 4964 (1974).
\bibitem{nightingale82} P. Nightingale, J. Appl. Phys. {\bf 53}, 7927 (1982).
\bibitem{beale86} P.D. Beale, Phys. Rev. B {\bf 33}, 1717 (1986).
\bibitem{jain80} A.K. Jain ans D.P. Landau, Phys. Rev. B {\bf 22}, 445 (1980).
\bibitem{landau81} D.P. Landau and R.H. Swendsen, Phys. Rev. Lett. {\bf 46}, 1437 (1981).
\bibitem{care93} C.M. Care, J. Phys. A {\bf 26}, 1481 (1993).
\bibitem{deserno97} M. Deserno, Phys. Rev. E {\bf 56}, 5204 (1997).
\bibitem{Blote95} H.W.J. Bl\"{o}te, E. Luijten, and J.R. Heringa, J. Phys. A {\bf 28}, 6289 (1995).
\bibitem{Blote04} Y. Deng and H.W.J. Bl\"{o}te, Phys. Rev. E {\bf 70}, 046111 (2004).
\bibitem{silva06} C.J. Silva, A.A. Caparica, and J.A. Plascak, Phys. Rev. E {\bf 73}, 036702 (2006).
\bibitem{metro53} N. Metropolis, A.W. Rosenbluth, M.N. Rosenbluth, A.H. Teller, and E. Teller, J. Chem. Phys. {\bf 21}, 1087 (1953).
\bibitem{Swendsen87} R.H. Swendsen and J.S. Wang, Phys. Rev. Lett. {\bf 58}, 86 (1987); U. Wolff,\emph{ibid.} {\bf 62}, 361 (1989).
\bibitem{Newman99} M.E.J Newman and G.T. Barkema, \emph{Monte Carlo Methods
in Statistical Physics} (Clarendon, Oxford, 1999).
\bibitem{LandBind00} D.P. Landau and K. Binder, \emph{Monte Carlo Simulations
in Statistical Physics} (Cambridge University Press, Cambridge,
2000).
\bibitem{bittner11} E. Bittner and W. Janke, arXiv: 1107.5640v1.
\bibitem{malakis12} A. Malakis, G. Gulpinar, Y. Karaaslan, T. Papakonstantinou, and G. Aslan, Phys. Rev. E {\bf 85}, 031146 (2012).
\bibitem{DSFish95} D.S. Fisher, Phys. Rev. B {\bf 51}, 6411 (1995).
\bibitem{Chay86} J.T. Chayes, L. Chayes, D.S. Fisher, and T. Spencer, Phys. Rev. Lett. {\bf 57}, 299 (1986); Comm. Math. Phys. {\bf 120}, 501 (1989).
\bibitem{Ludw90} A.W.W. Ludwig, Nucl. Phys. B {\bf 330}, 639 (1990).
\bibitem{Month09} C. Monthus, B. Berche, and C. Chatelain, J. Stat. Mech.: Theory Exp. (2009) P12002.
\bibitem{ferrenberg91} A.M. Ferrenberg and D.P. Landau, Phys. Rev. B {\bf 44}, 5081 (1991).
\bibitem{binder81} K. Binder, Z. Phys. B {\bf 43}, 119 (1981).
\bibitem{Guida98} R. Guida and J. Zinn-Justin, J. Phys. A {\bf 31}, 8103 (1998).
\bibitem{Hase10} M. Hasenbusch, Phys. Rev. B {\bf 82}, 174433 (2010).
\bibitem{Campo02} M. Campostrini, A. Pelissetto, P. Rossi, and E. Vicari, Phys. Rev. E {\bf 65}, 066127 (2002).
\bibitem{Bute02} P. Butera and M. Comi, Phys. Rev. B {\bf 65}, 144431 (2002); \emph{ibid.} {\bf 72}, 014442 (2005).
\end{thebibliography}
\end{document}